\documentstyle[preprint,aps]{revtex}
\begin{document}
\input{epsf}
\draft
\newfont{\form}{cmss10}
\newcommand{\e}{\varepsilon} 
\renewcommand{\b}{\beta} 
\newcommand{\unity}{1\kern-.65mm \mbox{\form l}}
\newcommand{\D}{D \raise0.5mm\hbox{\kern-2.0mm /}}
\newcommand{\A}{A \raise0.5mm\hbox{\kern-1.8mm /}}
\def\pmb#1{\leavevmode\setbox0=\hbox{$#1$}\kern-.025em\copy0\kern-\wd0
\kern-.05em\copy0\kern-\wd0\kern-.025em\raise.0433em\box0}

\def\D{\hbox{\hbox{${D}$}}\kern-1.9mm{\hbox{${/}$}}}
\def\kbar{\hbox{$k$}\kern-0.2truecm\hbox{$/$}}
\def\nbar{\hbox{$n$}\kern-0.23truecm\hbox{$/$}}
\def\pbar{\hbox{$p$}\kern-0.18truecm\hbox{$/$}}
\def\nhbar{\hbox{$\hat n$}\kern-0.23truecm\hbox{$/$}}
\newcommand{\dif}{\hspace{-1mm}{\rm d}}
\newcommand{\dil}[1]{{\rm Li}_2\left(#1\right)}
\newcommand{\diff}{{\rm d}} 

\title{$q\bar q$ interaction in light-cone gauge formulations of\\ 
Yang--Mills theory in 1+1 dimensions}
\author{A. Bassetto ($^*$)}
\address{CERN, Theory Division, CH-1211 Geneva 23, Switzerland\\
INFN, Sezione di Padova, Italy}
\author{D. Colferai}
\address{Dipartimento di Fisica ``G.Galilei", Via Marzolo 8, 35131
Padova, Italy}
\author{G. Nardelli}
\address{Dipartimento di Fisica , Universit\`a di Trento,
38050 Povo (Trento), Italy \\ INFN, Gruppo Collegato di Trento, Italy}
\maketitle
\begin{abstract}
A rectangular Wilson loop with sides parallel to space and time directions 
is perturbatively evaluated in two light-cone gauge formulations of 
Yang--Mills theory in 1+1 dimensions, with ``instantaneous'' and 
``causal'' interactions between static quarks.
In the instantaneous formulation we get Abelian-like exponentiation of 
the area in terms of $C_F$.
In the ``causal'' formulation the loop depends not only on the area,
but also on the dimensionless ratio $\beta = {L \over T}$, 
$2L$ and $2T$ being the
lengths of the rectangular sides. Besides it also exhibits dependence
on $C_A$. 
In the limit $T \to \infty$ the area law is recovered, but dependence on
$C_A$ survives. Consequences of these results are pointed out. 
\end{abstract} 

\noindent
($^*$) On leave of absence from
Dipartimento di Fisica ``G.Galilei", Via Marzolo 8 -- 35131
Padova, Italy.
\vfill\eject

\narrowtext

\section{Introduction}
\noindent

Starting from the pioneering work of
't Hooft \cite{Hoof}, pure Yang--Mills theory (YM) in the light-cone
gauge $A_-=0$ 
in two dimensions is considered a free theory, apart perhaps from topological
effects. This feature is at the 
root of the possibility of calculating the mesonic spectrum in the 
large-$N$ approximation \cite{Hoof,Cavi} when  dynamical  quarks are introduced. 

However the theory exhibits severe infrared (IR) divergences, which 
need to be regularized. In ref.\cite{Hoof} an explicit IR cutoff is advocated, 
which turns out to have no influence on the bound state equation; a
Cauchy principal value (CPV) prescription in handling such IR singularity
leads indeed to the same result \cite{Call}. On the other hand
this prescription emerges quite naturally if the theory is quantized at 
``equal $x^+$'', namely adopting the light-front $x^+=0$ as quantization
surface. 

Still difficulties in performing a 
Wick's rotation in the dynamical equations was pointed out
in ref.\cite{Wu}. In order to remedy such a situation, a causal 
prescription for the double pole in the
kernel was proposed, which is nothing but the one suggested
years later by Mandelstam and Leibbrandt (ML) \cite{ML}, 
when restricted to $1+1$ dimensions. This prescription
follows from equal-time quantization of the theory \cite{Bas5}.

Then
a quite different solution for the quark self-energy was obtained,
whereas the integral equation for $q\bar q$ bound states turned out
to be very difficult to solve. An approximate numerical treatment in the
massless quark case  led to a
quite different spectrum \cite{BSW}. A similar conclusion is reached
by a  recent analytical investigation for different values of the
input parameters \cite{BNS}.

In view of the above-mentioned controversial results and of the
fact that ``pure'' YM theory does not immediately look free in Feynman gauge,
a test of gauge invariance was performed 
in ref.\cite{Bas1} by calculating at ${\cal 
O}(g^4)$, both in Feynman and in light-cone gauge with ML prescription,
a rectangular Wilson loop with 
light-like sides, directed along the vectors $n_\mu = (T, - T)$, 
$n^*_\mu = (L, L)$ and parametrized according to the equations:

\begin{eqnarray}
\label{uno}
C_1 &:& x^\mu (t) = n^{* \mu} t,\nonumber\\
C_2 &:& x^\mu (t) = n^{* \mu} + n^\mu t,\nonumber\\
C_3 &:& x^\mu (t) = n^\mu + n^{* \mu}( 1-t), \nonumber\\
C_4 &:& x^\mu (t) = n^\mu (1 - t), \qquad 0 \leq t \leq 1. 
\end{eqnarray}

This contour has been considered in refs.\cite{Korc,Bas2} for
an analogous test of gauge invariance in 1+3 dimensions. Its 
light-like character forces a Minkowski treatment.

In order to perform the test, dimensional regularization was used;
the Feynman option is indeed not viable at strictly 1+1 dimensions,
as the usual free vector propagator is not a tempered distribution.

The following unexpected results were obtained.

The ${\cal O}(g^4)$ perturbative loop expression in $d= 1+(D - 1)$
dimensions is finite in the limit $D\to 2$. The results in the two gauges
coincide, as required by gauge invariance. They exhibit dependence
on $C_A$, the Casimir constant of the adjoint representation.

This dependence, when looked at in the light-cone gauge calculation,
comes from non-planar diagrams with the colour factor $C_F(C_F - C_A/2)$,
$C_F$ being the Casimir constant of the fundamental representation.
Besides, there is a genuine contribution proportional to $C_F C_A$
coming from the one-loop correction to the vector propagator. This is
surprising at first sight, as in strictly 1+1 dimensions the
triple vector vertex vanishes in axial gauges. What happens is that
transverse degrees of freedom, although coupled with a
vanishing strength at $D=2$, produce finite contributions when
matching with the self-energy loop singularity precisely at $D=2$,
eventually producing a finite result. Such a
dimensional anomaly-type phenomenon could not appear in a strictly
1+1 dimensional calculation, which would only lead to the (smooth)
non-planar diagram result. We stress  that this ``anomalous''
contribution is essential to get agreement with the Feynman gauge
calculation, in other words with gauge invariance.
We notice that no ambiguity affects our light-cone gauge results, 
which do not involve infinities; in addition the discrepancy 
cannot be accounted for by a simple redefinition of the coupling, 
that would also, while unjustified on general grounds, turn out 
to be dependent on the area of the loop.

\smallskip

We are led to the following conclusion: either the theory has a basic
discontinuity at $D=2$ or a light-cone gauge Wilson loop
calculation
with the ML prescription in strictly 1+1 dimensions, where the triple
vector vertex is zero, is at odds with gauge invariance, as the above-mentioned
``anomalous''contribution would be missing.

\smallskip

In order to make the argument complete, we recall that a calculation
of the same Wilson loop in strictly 1+1 dimension in light-cone
gauge with CPV prescription for the singularity produces a
vanishing contribution from non-planar graphs. Only planar diagrams
survive, leading to Abelian-like contributions depending only
on $C_F$, which can be resummed to all orders in the perturbative
expansion to get the expected exponentiation of the area.
This result, which is the usual one found in the literature,
although quite transparent, {\it does not coincide}
with the Feynman gauge result in the limit $D\to 2$.
Again we do not see any sensible way to reconcile the two results.
The test cannot be generalized to $D\ne 2$ dimensions as
CPV prescription leads to inconsistency with
power counting in Feynman diagrams in this case\cite{Bas3}.

\smallskip
In order to clarify whether there is indeed a pathology in the light-cone
gauge formulation with ML prescription in strictly 1+1 dimensions, one can
try to study the potential $V(2L)$ between a 
``static" quark--antiquark pair in the fundamental representation, 
separated by a distance $2L$. Then a different
Wilson loop is to be calculated, {\it i.e.} a rectangular loop with
one side along the space direction and one side along the time direction,
of length $2L$ and $2T$ respectively. Eventually the limit $T \to \infty$
is to be considered: the potential $V(2L)$  between the
quark and the antiquark is indeed 
related to the value of the corresponding Wilson loop
amplitude ${\cal W}(L,T)$ through the well-known 
formula\cite{Pol,Kog,Fis,nota1}
\begin{equation}
\lim_{T\to\infty}{\cal W}(L,T)= e^{-2i T V(2L)}\ .
\label{potential}
\end{equation}

The crucial point to notice here is that dependence on the Casimir
constant $C_A$ should cancel at the leading order in any
coefficient of a perturbative expansion of the potential with
respect to coupling constant. This criterion has often been used
as a check of gauge invariance.

\smallskip

These are just  motivation and  content of the present paper.

In Sect. II we recall definitions and general properties of the Wilson
loop we are going to evaluate.

In Sect. III we  present our results. In the CPV
case, due to its essentially Abelian nature, the loop can be exactly
evaluated, the so-called area law is recovered, thus providing a
linear potential between the quark and the antiquark. In particular
full (Abelian-like) exponentiation in terms of only $C_F$ will occur.

The corresponding calculation in the ML case
develops genuine non-Abelian terms proportional to $C_A$; 
thus, contrary to the previous case, the loop
interaction feels the non-Abelian nature of the theory.
Higher-order perturbative contributions cannot be simply
computed; we limit ourselves to a perturbative ${\cal O}(g^4) $
calculation, which nevertheless turns out to be sufficient for our
purposes. The result will be that not only does the loop not
obey a simple Abelian exponentiation, but even the  ``area law'' is lost for
finite values of $T$ and $L$.
In the limit $T \to \infty$ the area law is recovered, but dependence on
$C_A$ survives. 

\smallskip

Our conclusions are drawn in Sect. IV and technical details 
are given in the Appendices.

\section{General Considerations} 
\noindent
In $1+1$ dimensions, we consider the usual ``gauge fixed'' Yang-Mills
Lagrangian,
\begin{equation}
{\cal L} = -{1\over 4} F^a_{\mu \nu} F^{a\mu \nu} -\lambda^a (nA^a)\  ,
\label{lagrangian}
\end{equation}
where $\lambda^a$ are Lagrange multipliers enforcing the light-cone gauge
condition $n^\mu A_\mu^a= A_-^a=0$, $n_\mu=(1/\sqrt{2})(1,1)$ being a constant
(gauge) vector.
Without loss of generality, we consider $SU(N)$ as gauge group, so that
the field strength in (\ref{lagrangian}) is  defined as $F_{\mu \nu}^a=
\partial_\mu A_\nu^a - \partial_\nu A_\mu^a + g f^{abc} A_\mu^b A_\nu^c$,
$f^{abc}$ being the structure constants of $SU(N)$. 
Once the gauge condition $A_-^a=0$ has been taken into account, the interaction
term in the field strength vanishes, so that the theory is manifestly free and
the free propagator turns out to be the complete two-point Green function.
In turn, the prescription for handling the poles of the Green function is fixed
by the quantization procedure.

Equal $x^+$ quantization entails the expression for the free propagator:
\begin{equation}
D_{++}^{(CPV)ab}(x)=D_{++}^{(CPV)}(x)\delta^{ab} = 
- {i\delta^{ab}\over (2\pi)^2} \int d^2k\,
e^{ikx} {\partial\over \partial k_-} CPV\left({1\over k_-}\right)=-
{i\delta^{ab}\over 2} |x^-|\delta(x^+)\ , \label{propcpv}
\end{equation}
whereas equal time quantization leads to
\begin{equation}
D_{++}^{(ML)ab}(x) =D_{++}^{(ML)}(x) \delta^{ab} = 
{i \delta^{ab}\over \pi^2}\int d^2k
\, e^{ikx} {k_+^2\over (k^2 + i \epsilon)^2}= {\delta^{ab}\over \pi}
{(x^-)^2\over (-x^2 + i\epsilon)}\  \label{propml}
\end{equation}
for the same quantity, with CPV and ML respectively.

\smallskip

We consider the closed path $\gamma$  
parametrized by the following  four segments
$\gamma_i$,
\begin{eqnarray}
\gamma_1 &:& \gamma_1^\mu (s) = (sT, L)\ ,\nonumber\\
\gamma_2 &:& \gamma_2^\mu (s) = (T,-sL)\ ,\nonumber\\
\gamma_3 &:& \gamma_3^\mu (s) = (-sT, -L)\ , \nonumber\\
\gamma_4 &:& \gamma_4^\mu (s) = (-T, sL)\ , \ \ \qquad -1 \leq s \leq 1. 
\label{path}
\end{eqnarray}
describing a  (counterclockwise-oriented) rectangle
 centred at the origin of the plane ($x^1,x^0$),
with length sides $(2L,2T)$, respectively (see Fig. \ref{fig1}).
Then, for the definition of the Wilson loop around $\gamma$ we shall adopt the
standard one, given by the following  vacuum to vacuum amplitude
\begin{equation}
{\cal W}_\gamma (L,T) = {1\over N} \langle 0| {\rm Tr}\left[ {\cal T}{\cal P} 
{\rm exp} \left( ig \oint_\gamma dx^\mu \ A^a_\mu (x) T^a \right)\right]
|0\rangle \ \ , \label{wilson}
\end{equation}
where ${\cal T}$ orders gauge fields in time and ${\cal P}$ orders generators
$T^a$ of the gauge group $SU(N)$ along the closed integration path $\gamma$. 
The perturbative expansion of the Wilson loop (\ref{wilson}) looks like
\begin{equation}
{\cal W}_\gamma (L,T) = 1 + {1\over N}\sum_{n=2}^\infty (ig)^n \oint_\gamma
dx_1^{\mu_1} \cdots \oint_\gamma dx_n^{\mu_n}\theta( x_1 >\cdots >x_n )
{\rm Tr} [ G_{\mu_1 \cdots \mu_n} (x_1,\cdots ,x_n)]\ ,
\label{wilpert}
\end{equation}
where $ G_{\mu_1 \cdots \mu_n} (x_1,\cdots ,x_n)$ is the Lie algebra valued
$n$-point Green function, in which further dependence on the coupling
constant is usually buried; the Heavyside $\theta$-functions order the points 
$x_1,\cdots ,x_n$ along the integration path $\gamma$.

It is easy to show that the perturbative expansion of ${\cal W}_\gamma$ is an
even power series in the coupling constant, so that we can write
\begin{equation}
{\cal W}_\gamma (L,T)= 1+g^2 {\cal W}_2 + g^4 {\cal W}_4 + {\cal O}(g^6)\ .
\label{pert}
\end{equation}

The fact that we are in strictly 1+1 dimensions greatly simplifies the
perturbative expansion as the complete Green functions are just products
of free propagators.

An explicit evaluation of the function  ${\cal W}_2$ in eq. (\ref{pert})
gives the diagrams contributing to the loop with a single exchange (i.e. one
propagator), namely
\begin{equation}
{\cal W}_2= - {1\over 2} C_F \oint \oint D_{\mu\nu} (x-y) dx^\mu dy^\nu \ .
\label{w2}
\end{equation}

Concerning ${\cal W}_4$, a straightforward calculation gives
\begin{equation}
{\cal W}_4={1\over 8N}\oint\oint\oint\oint{\rm Tr}[{\cal P}
 (T^a_xT^a_yT^b_z T^b_w)]
  D_{\mu\nu}(x-y)D_{\rho\sigma}(z-w)\;
 dx^{\mu}dy^{\nu}dz^{\rho}dw^{\sigma}\;\; ,
\label{w4}
\end{equation}
where subscripts in the matrices have been introduced to specify their ordering.
>From eq. (\ref{w4}), the diagrams with two-gluons exchanges contributing to the
order $g^4$ in the perturbative expansion of the Wilson loop fall into two
distinct classes, depending on the topology of the diagrams:
\begin{enumerate} 
\item {\it Uncrossed diagrams}: if the pairs $(x,y)$ and $(z,w)$
are contiguous around the loop the two propagators do not cross (see Fig.
\ref{fig2}) and the trace in (\ref{w4}) gives ${\rm Tr} [T^aT^aT^bT^b] = N
C_F^2$. 
\item {\it Crossed diagrams}: if the pairs $(x,y)$ and $(z,w)$
are  not contiguous around the loop the two propagators do cross (see
Fig. \ref{fig2}) and the trace in (\ref{w4}) gives ${\rm Tr} [T^aT^bT^aT^b] =
{\rm Tr}[ T^a (T^a T^b + [T^b,T^a])T^b]= N (C_F^2 - (1/2) C_A C_F)$, $C_A$ being
the Casimir constant of the adjoint representation defined by
$f^{abc}f^{d bc}= C_A \delta^{ad}$.
\end{enumerate}
We see that  the $C_F^2$ term is present in both types of diagrams and with
the same coefficient. This term is usually denoted  ``Abelian term'': were the
theory Abelian, only such $C_F^2$ terms would contribute to the loop. On the
other hand, the $C_FC_A$ term is present only in crossed diagrams, and is
typical of non-Abelian theories.

Thus, we can decompose  ${\cal W}_4$ as the sum of an Abelian and a 
non-Abelian part,
\begin{equation}
{\cal W}_4={\cal W}_4^{ab} +{\cal W}_4^{na}\ \ .
\end{equation}
 Moreover,  the Abelian part is simply half of the square of the 
order-$g^2$ term, i.e.
\begin{eqnarray}
{\cal W}_4^{ab} &= & {1\over 8} C_F^2 
\oint\oint\oint\oint
       D_{\mu\nu}(x-y)D_{\rho\sigma}(z-w)\;
        dx^{\mu}dy^{\nu}dz^{\rho}dw^{\sigma}\nonumber\\
&=& {1\over 2} \left( - {1\over 2} C_F \oint \oint D_{\mu\nu} (x-y) 
dx^\mu dy^\nu \right)^2 \ \ .\label{w4ab}
\end{eqnarray} 
Equation (\ref{w4ab}) is just a particular case of a more general theorem due to
Frenkel and Taylor \cite{Fre}, which proves that
the only relevant 
terms in the perturbative expansion of the loop are the so-called ``maximally
non-Abelian'' ones; at ${\cal O}(g^4)$ those terms are just proportional to
$C_FC_A$. In turn they only come from crossed diagrams in our case, so that an
analysis of just crossed diagrams is enough to get the complete ${\cal
O}(g^4)$ expansion of the Wilson loop, once the ${\cal O}(g^2)$ term is known.

All the Abelian  terms
(depending only on $C_F$) in
the perturbative expansion of the Wilson loop sum up to reproduce the Abelian
exponential
\begin{equation}
{\cal W}_\gamma^{ab} (L,T) = {\rm exp}\left( - {1\over 2} C_F g^2
\oint\oint D_{\mu \nu}(x-y) dx^\mu dy^\nu \right) \  . \label{abexp}
\end{equation}

\section{Wilson Loop Results} 
\noindent 
We have now to calculate  loop
integrals of the type given in eqs. (\ref{w2}) and (\ref{w4}). In view of 
the parametrization  (\ref{path}), it is convenient to decompose loop integrals
as sums of integrals  over the segments $\gamma_i$, and to this purpose we
define
\begin{equation}
E_{ij}(s,t) = D_{\mu \nu}\bigl[ \gamma_i(s) -\gamma_j(t) \bigr]
\dot\gamma_i^\mu(s) \dot\gamma_j^\nu(t) \ , \qquad i,j=1,\dots ,4\ \, , \label{e}
\end{equation}
where the dot denotes the 
derivative with respect to the variable parametrizing the
segment. In this way, each diagram can be written as integrals of products of
functions of the type (\ref{e}).  Each graph will be labelled by  a set of pairs
$(i,j)$, each pair denoting a gluon propagator joining the segments $\gamma_i$
and $\gamma_j$. 
Obviously, depending on whether we perform the calculation 
in the CPV formulation or in the ML one, we shall use in
eq. (\ref{e}) the propagators (\ref{propcpv}) or (\ref{propml}), respectively.
Due to the symmetric choice of the contour $\gamma$ and to the  fact that
propagators (both in the CPV and in the ML case) are even functions, i.e. $D_{\mu
\nu}(x)=D_{\mu \nu}(-x)$, we have the following identities that halve the
number of diagrams to be evaluated:
\begin{eqnarray}
E_{ij}(s,t) &=& E_{ji}(t,s)\ ,\nonumber\\
E_{12}(s,t) &=& E_{34}(s,t)\ ,\nonumber\\
E_{23}(s,t) &=& E_{41}(s,t)\ ,\nonumber\\
E_{11}(s,t) &=& E_{33}(s,t)\ ,\nonumber\\
E_{22}(s,t) &=& E_{44}(s,t)\ .
\label{sym}
\end{eqnarray}

\subsection{Calculation in the CPV Formulation}
\noindent
We shall begin with the ${\cal O} (g^2)$ contribution. Since the integrations
are elementary in this case, we shall only report the final result:
 from  eq. 
(\ref{w2}), using eqs. (\ref{path}), (\ref{propcpv}), (\ref{e}) and
(\ref{sym}) one can easily get
\begin{equation}
{\cal W}_2^{CPV} = - {1\over 2} C_F \sum_{i,j=1}^4 \int_{-1}^1 ds\int_{-1}^1
dt\ E_{ij}^{CPV} (s,t) =-{i\over 2} C_F A\ ,
\label{g2cpv}
\end{equation}
$A=4LT$ being the area of the loop $\gamma$.

Once the ${\cal O}(g^2)$ term is known, simple arguments permit to evaluate 
any order in the perturbative expansion, so that in this case
the loop can be exactly obtained. Let us indeed consider the ${\cal
O}(g^4)$ term. As explained in the previous section, only the ``genuine'' 
non-Abelian part needs to be evaluated, i.e. the crossed diagrams containing the
factor $C_FC_A$, as the Abelian part is already given by eq. (\ref{w4ab}), which
in this case leads to 
\begin{equation}
{\cal W}_4^{CPV,ab}= -{1\over 8}C_F^2 A^2\ . 
\end{equation}
However, due to the contact nature of the propagator in the $CPV$ case, all the
crossed diagrams trivially vanish so that 
${\cal W}_4^{CPV,na}=0$: the $\delta(x^+)$ term in the propagator only 
tolerates
diagrams with parallel propagators, which 
therefore cannot cross, both in the Abelian and in the non-Abelian case. 
Obviously, this
argument holds at any order in the perturbative expansion, so that only {\it
Abelian} terms contribute and the sum of all of
them, due to the Abelian exponentiation theorem, reproduces the exponential
\begin{equation}
{\cal W}_\gamma^{CPV} = {\rm exp}\left(-{i\over 2} g^2 C_F A\right)\ .
\label{wlcpv}
\end{equation}
A detailed discussion of this point can be found in Appendix A.

Consequently, from eq. (\ref{potential}),  null plane light-cone quantization
provides a linear confining potential for a quark--antiquark pair, with string
tension $\sigma= g^2 C_F/2$. 
This is the very same result one would have obtained in an Abelian theory.
In this sense null plane light-cone gauge quantization provides a ``free''
theory: the Wilson loop does not feel the  non-Abelian colour structure 
of the theory. This result, obtained in a Minkowskian framework, 
is in agreement with  analogous
Euclidean calculations\cite{Bra}.

\subsection{Calculation in the ML Formulation }
\noindent
Unfortunately, in this case the calculations are much more complicated. We
begin with the ${\cal O}(g^2)$ terms. Following the notation introduced in
the previous section, ${\cal W}_2^{ML}$ can be written as the sum of 16 diagrams
\begin{eqnarray}
{\cal W}_2^{ML}&=& - {1\over 2} C_F \sum_{i,j=1}^4 \int_{-1}^1 ds\int_{-1}^1 dt
E_{ij}(s,t)\nonumber\\
 &\equiv &- {1\over 2} C_F \sum_{i,j=1}^4 C_{ij}\ .\label{m}
\end{eqnarray}
Thanks to the symmetry properties (\ref{sym}), only 6 of them are independent,
and an explicit evaluation gives

\begin{eqnarray}
C_{11}&=&C_{33}=\frac{L^2}{\pi}\;\left(-\frac{1}{\b^2}\right)\;\;\nonumber\\
C_{22}&=&C_{44}=\frac{L^2}{\pi}\;\;\nonumber\\
C_{12}&=&C_{21}=C_{34}=C_{43}=\frac{L^2}{\pi}\left[i\pi-\ln(\b)+
             \left(1-\frac{1}{\b^2}\right)\ln(1-\b)\right]\;\;\nonumber\\
C_{14}&=&C_{41}=C_{23}=C_{32}=\frac{L^2}{\pi}\left[-\ln(\b)+\left(
           1-\frac{1}{\b^2}\right)\ln(1+\b)\right]\;\;\nonumber\\
C_{13}&=&C_{31}=\frac{L^2}{\pi}\bigg[\frac{1}{\b^2}+\left(
     \frac{2}{\b}-2\right)i\pi+4\ln(\b)-
        \left(\frac{2}{\b}+2\right)\ln(1+\b)+\left(
            \frac{2}{\b}-2\right)\ln(1-\b)\bigg]\;\;\nonumber\\
C_{24}&=&C_{42}=\frac{L^2}{\pi}\left[-1+\left(\frac{2}{\b^2}
            +\frac{2}{\b}\right)
            \ln(1+\b)+\left(\frac{2}{\b^2}-\frac{2}{\b}\right)
            \ln(1-\b)\right]\;\;\ \label{cij}
\end{eqnarray}
Summing up all the coefficients (\ref{cij}) as in (\ref{m}) one gets that the
second-order calculation is in agreement with the CPV case, i.e.
\begin{equation}
{\cal W}_2^{ML}= {\cal W}_2^{CPV}= -{i\over 2} C_FA\ . 
\label{mlo2}
\end{equation}
However, as often happens in Wilson loop calculations, an ${\cal O}(g^2)$ 
computation is too weak a probe to check consistency and gauge invariance. 
Thus, we have to consider the ${\cal O}(g^4)$ terms. Again, only ``crossed
diagrams'' (maximally non-Abelian ones)  need to be evaluated

\begin{eqnarray}
{\cal W}_4^{ML, na}& =& - {1\over 2} C_AC_F
\sum_{i,j,k,l}{}^{'} \int ds\int dt\int du\int dv \ 
E_{ij}(s,t)E_{kl}(u,v)\nonumber\\ &\equiv&- {1\over 2} C_AC_F
\sum_{i,j,k,l}{}^{'} C_{(ij)(kl)}\ ,
\label{doppi}
\end{eqnarray}
where the primes mean that we have to sum only over
crossed propagators configurations and over topologically inequivalent
contributions, as carefully explained in the following;
we have not specified the integration
extrema as they depend on the particular type of crossed diagram we are
considering (the extrema
must be chosen in such a way that propagators remain crossed). 

The last equality in eq. (\ref{doppi}) defines
the general diagram $C_{(ij)(kl)}$: it is a  diagram with two {\it crossed}
propagators joining the sides $(ij)$ and $(kl)$ of the contour (\ref{path}).
In Fig. \ref{fig3} a few examples of diagrams are drawn to get the reader
acquainted with the
notation.
 The first of eq.
(\ref{sym})  permits to select just 35 types of topologically distinct
crossed diagrams, and to multiply each representative by a factor $8$, which is
the number of permutations of the points $(x,w,y,z)$ leaving the propagators 
$D_{\mu\nu}[x(s)-y(t)]D_{\sigma\rho}[w(u)-z(v)]$ crossed. The remaining symmetry
relations (\ref{sym}) further lower the number to 19. 
As a matter of fact, although topologically inequivalent, from eq. (\ref{sym})
it is easy to get  
\begin{eqnarray}
C_{(11)(11)}&= C_{(33)(33)}\ \ , \qquad C_{(22)(22)}=& C_{(44)(44)}\
\ ,\nonumber\\
C_{(11)(13)}&= C_{(33)(13)}\ \ , \qquad C_{(22)(24)}=& C_{(44)(24)}\
\ ,\nonumber\\
C_{(11)(12)}&= C_{(33)(34)}\ \ , \qquad C_{(22)(23)}=& C_{(44)(14)}\
\ ,\nonumber\\
C_{(11)(14)}&= C_{(33)(23)}\ \ , \qquad C_{(22)(12)}=& C_{(44)(34)}\
\ ,\nonumber\\
C_{(13)(12)}&= C_{(13)(34)}\ \ , \qquad C_{(24)(23)}=& C_{(24)(14)}\
\ ,\nonumber\\
C_{(13)(14)}&= C_{(13)(23)}\ \ , \qquad C_{(24)(12)}=& C_{(24)(34)}\
\ ,\nonumber\\
C_{(12)(14)}&= C_{(23)(34)}\ \ , \qquad C_{(12)(23)}=& C_{(14)(34)}\
\ ,\nonumber\\
C_{(12)(12)}&= C_{(34)(34)}\ \ , \qquad C_{(23)(23)}=& C_{(14)(14)}\ \ ,
\label{relations}
\end{eqnarray}
which are the 16 
relations needed to lower the number of diagrams to be evaluated
from 35 to 19. Besides the 32 diagrams quoted in eq. (\ref{relations}), there
are three other 
crossed diagrams that do not possess any apparent symmetry relation
with other diagrams: $C_{(13)(13)}, \ C_{(24)(24)}$ and $C_{(13)(24)}$ (see
Fig. \ref{fig4}), so that the number of topologically inequivalent crossed
diagrams is indeed 35.

The
calculation of the 19 independent  diagrams needed is lengthy and  not trivial.
The details of such calculation are fully
reported in the Appendix B. Each diagram depends not only on the area
$A=4LT$ of the loop, but also on the dimensionless ratio $\beta = L/T$ through
complicated functions involving powers, logarithms and dilogarithm functions,
denoted by ${\rm Li}_2(z)$.
 Since we shall be interested in the large-$T$
behaviour, we have always considered  the region $\beta<1$, and in
the final result we have performed all the analytical continuations of the
dilogarithm in such a way that the point $\beta=0$ corresponds to the
vanishing argument of ${\rm Li}_2(z)$, in order to have easily expandible
expressions (see Appendix B for details).

Adding all the contributions as in eq. (\ref{doppi}) we eventually arrive at the
following result for the non-Abelian part of the ${\cal O}(g^4)$ contributions:
\begin{eqnarray}
 {\cal W}_4^{ML,na}
&=&\frac{C_AC_FA^2}{32\pi^2}\,\Bigg\{\frac{2}{3}\pi^2
               -\frac{2}{3\b}\ln^2(1+\b)
               -\frac{1}{6\b^2}[1-\b]^4\ln^2(1-\b)
-\frac{1}{3\b^2}[1-\b]^4\times \Bigg.\nonumber\\ &&\Bigg.\left[ { \over
}\dil{\b}
+\dil{-\frac{\b}{1-\b}}
               \right]-\frac{4}{3\b}\left[\dil{-\b}+\dil{\frac{\b}{1+\b}}\right]
               \Bigg\}\;\; .
\label{wmlg4}
\end{eqnarray}
Several important consequences can be drawn from eq. (\ref{wmlg4}):
\begin{enumerate} 
\item The sum of all non-Abelian terms, proportional to $C_FC_A$, does not
vanish. This fact prevents any possible agreement with the CPV
formulation, where the result was a simple Abelian exponentiation, see eq.
(\ref{wlcpv}).
\item The result (\ref{wmlg4}) does not depend only on the area $A$ of the
loop, but also on the ratio $\beta=L/T$. 
It is remarkable, and perhaps not incidental, that for $\beta \to 0$ (large $T$)
a pure area dependence is recovered, i.e.
\begin{equation}
\lim_{\b\to0}{\cal W}_4^{ML,na}=\frac{1}{48} C_AC_F\;A^2\;\;. 
\label{limit}
\end{equation}
Nevertheless, since the above limit does not vanish, even at
 large $T$  ${\cal W}_\gamma^{ML}$ fails to have an  Abelian exponential
behavior.

\item A little thought is enough to realize that, also in the large-$T$
limit, the perturbative series ${\cal W}_\gamma^{ML} $ cannot
sum to a {\it phase factor}, 
even taking into account possible  extra non-Abelian terms in the argument of
the exponent. As a matter of fact, following  
\cite{Gat}, from (\ref{limit}) and 
(\ref{mlo2}) one  concludes that 
\begin{equation}
\lim_{T\to \infty} {\cal W}_\gamma^{ML}= {\rm exp}\left\{-{i\over 2}C_F  g^2 A
+\frac{1}{48}C_AC_F g^4 A^2 \right\} + {\cal
O} (g^6)\ . \label{exp}
\end{equation}
\end{enumerate}

As the calculation ${\cal O}(g^4)$ is really very heavy, we have performed a 
consistency check of its accuracy. We have indeed independently
computed the contribution from {\it uncrossed} graphs
${\cal O}(g^4)$ (which only involve $C_F^2$) and then we have summed it
to the expression for the corresponding $C_F^2$ from the {\it crossed}
graphs, which has twice the weight in eq. (\ref{limit}); in so doing
the full ${\cal O}(g^4)$ Abelian result has been correctly recovered.

\section{Discussion} 
\noindent 
We have explored in $1+1$ dimensions two inequivalent formulations of 
Yang--Mills theory, within the same gauge choice (the light-cone gauge $A_-=0$).
One of them quantizes the theory on the null plane $x^+=0$:
there are no propagating degrees of freedom, the only
non-vanishing component of the gauge potential $A_+$ not being a dynamical
variable, but just providing  a non-local Coulomb-type  force
between fermions (the ``propagator'' (\ref{propcpv})).
This formulation of  $1+1$ dimensional Yang--Mills theory has quite
often been considered in the literature and leads to rising Regge trajectories 
for quark--antiquark mesons, when dynamical fermions are included\cite{Hoof}.

In this case, due to the instantaneous nature of the interaction,
the exact  Wilson loop vacuum to vacuum expectation
value can be calculated: the
perturbative series can be summed to the exponential (\ref{wlcpv}), as
usually found in the literature.
Static quarks confine into  mesons   through
an attractive   linear  potential with string tension $\sigma =g^2 C_F/2$.
This is due to the essentially Abelian nature of the theory: confinement 
emerges in the same way as in $1+1$ dimensional
electrodynamics.
On the other hand
this formulation seems to exist
only in strictly  $1+1$ dimensions. As a matter of fact, in higher dimensions,
inconsistencies arise, at least in perturbative treatments,
which make the formulation 
unacceptable \cite{Bas3}. Moreover it cannot be viewed
as the $D\to 2$ limit of the theory in higher dimensions\cite{Bas1}.

Then we have considered the equal-time
light-cone gauge formulation that is suggested from higher dimensions.
It has indeed been shown that it provides the correct
way of handling Yang--Mills theory, in full agreement with 
Feynman gauge results \cite{Bas2}.
This formulation entails 
unphysical states that can be expunged from the ``physical''
Hilbert space, but are nevertheless necessary to obtain the causal form 
for the
propagator. This form unfortunately prevents a complete evaluation of 
the Wilson loop.

Consequently we have only performed a perturbative ${\cal
O}(g^4)$ calculation, which obviously is not sufficient to get information
on the $q\bar q$ potential, but still allows a non-trivial
comparison with the previous formulation.
For finite size $T$ and $L$ of the loop, the Wilson loop amplitude depends
not only on the area, but also on the dimensionless ratio $\beta=L/T$ through a
complicated factor involving the dilogarithm function ${\rm Li}_2(z)$, eq.   
(\ref{wmlg4}). In the limit $T\to \infty$ the area law is recovered, but
dependence on $C_A$ survives and the perturbative series
cannot exponentiate to a pure phase factor, even in this limit.

There is no way of reconciling this result with the one previously
found in the theory
quantized on the null plane, even in the large-$N$ limit. As a matter of fact
the dependence on $C_A$ occurs in this case only in the combination
$C_F-C_A/2$, which vanishes in the large-$N$ limit leaving a pure $C_F^2$
dependence; unfortunately, however, the coefficient of such a dependence,
which is due only to uncrossed graph configurations, cannot match
the Abelian-like one required by exponentiation.

\smallskip

The persistence of the dependence on $C_A$ in the leading coefficient of the
${\cal O}(g^4)$ expansion of a Wilson loop at large $T$
has always been interpreted as a pathology \cite{Bas3}. It means that
the ``causal'' version of light-cone gauge in strictly 1+1 dimensions
is sick, at least in perturbative calculations.

We have no explanation for such a phenomenon, which is
peculiar to the non-Abelian case (in QED both formulations lead to the
same result) and is at odds with all analogous calculations in higher
dimensions $d=1+(D-1)$, where the ``causal'' formulation seems to be the only
acceptable one, even in the limit $D\to 2$.

We only recall that in strictly 1+1 dimensions the term arising from 
the self-energy correction to the vector propagator is missing owing to the
vanishing of the triple vector vertex in light-cone gauge. Such a term
is present at $D>2$ instead and does not vanish in the limit $D\to 2$
\cite{Bas1}, thereby providing an ``anomaly''-type discontinuity, which
however is not sufficient to cancel the non-Abelian term.
We might expect an analogous behaviour from graphs with three vector lines
attached to the loop (``spider'' diagrams; examples are shown in Fig.5).

In dimensions higher than 2, a calculation using the CPV prescription,
besides violating causality and being extremely cumbersome, would not be 
reliable. It is indeed well known that already the gluon self-energy
at ${\cal O}(g^2)$, obtained by this regularization, is inconsistent
since it exhibits peculiar singularities which do not find cancellation
in the corresponding scalar and spinor contributions in SUSY N=4
\cite{CAPPER}. Those singularities do not appear either in Feynman
or in light-cone gauge with the ML prescription, where the finiteness
of SUSY N=4 can easily be proven \cite{DALBOSCO}.

In ref.\cite{Bas2} the coincidence of a rectangular Wilson loop result
in the $(x^+,x^-)$ coordinates was proven in $D$-dimensions at ${\cal O}
(g^4)$ using Feynman and light-cone gauge with ML prescription, respectively.

We expect that such a coincidence should persist also for an analogous
calculation in $D$-dimensions for a rectangular 
Wilson loop in the coordinates $(t,x^3)$. Then exponentiation should be
checked. The $C_A$ dependence should
disappear 
from the leading coefficient in the limit $T\to \infty$ and the limits 
$T\to \infty$ and $D\to 2$ should not be interchanged.

In conclusion we speculate that, in dimensions higher than 2,
exponentiation would occur in light-cone gauge with a causal formulation
(the only sensible one), whereas, in strictly 1+1 dimensions, it would
occur , although trivially, in the theory with a ``contact'' potential.
However, the latter theory cannot be reached by any 
continuous limiting procedure
from higher dimensions.

\acknowledgements
\noindent
We thank L. Griguolo for many useful discussions.

\vskip 1truecm
\centerline{\bf ERRATUM}

\noindent
In eq. (25) one can easily check, using eqs. (B24) and (B25),
that the dependence on the dimensionless ratio $\beta=L/T$
{\it exactly} cancels, leading to
a pure area behaviour for any value of $T$:

\begin{equation}
W_4^{ML,na} = {{C_A C_F A^2}\over {48}}.
\end{equation}

The unfortunate oversight of this cancellation led us to the
conclusion that the Wilson loop under consideration would
exhibit {\it at finite T}, besides the expected area dependence,
also a dependence on $\beta$,
which would cancel only in the limit $T \to \infty$.

This conclusion is wrong and should be taken back (see 
M. Staudacher and W. Krauth, hep-th/9709101).
However all other results are nicely confirmed, in particular the
coexistence of two physically different formulations at D=2 and
the dependence on $C_A$ in the leading coefficient of the ${\cal O}
(g^4)$ expansion of the loop at large $T$ in the Wu-Mandelstam-Leibbrandt
(WML) formulation.

In our opinion this dependence is a crucial feature and still puts 't Hooft's
and WML formulations on a different footing, as discussed in our
Sect. IV.

\vfill\eject

\appendix
\section{}

\noindent
In order to understand why crossed diagrams cannot contribute in the
CPV case, we first exhibit the quantites $E^{CPV}_{ij}(s,t)$.
Only two of them are independent, thanks to eq. (\ref{sym}), and
different from zero: 
\begin{equation}
E^{CPV}_{12}(s,t)={{iL^2}\over 2}(1+t)\delta(1-s-\beta (1+t)),
\end{equation}
and 
\begin{equation}
E^{CPV}_{13}(s,t)={{iL^2}\over {\beta}}\delta(s+t+2\beta),
\end{equation}
(we are considering the case $\beta<1$).

Let us look at the first diagram in Fig.4; its contribution would be
$C^{CPV}_{(13)(13)}$, according to the notation developed in eq. 
(\ref{doppi}). In this case the integration domain would be
constrained by the product $\delta(s+t+2\beta) \ \delta(u+v+2\beta)$,
with the conditions $t>u$ and $s>v$, to produce the crossing.
These conditions clearly cannot be fulfilled.

Another independent possibility would be $C^{CPV}_{(12)(13)}$.
The constraint now would be given by 
$\delta(1-s-\beta (1+t)) \ \delta(u+v+2\beta)$
with the crossing condition $u>s$, which is clearly impossible.

Finally $C^{CPV}_{(12)(12)}$ would be affected by the constraint
$\delta(1-s-\beta (1+t)) \ \delta(1-u-\beta(1+v))$ with the
conditions $s>u$ and $t>v$, which again are clearly impossible.

In higher orders the argument can be repeated considering the
propagators pairwise. On the other hand 
the conclusion on the vanishing of crossed diagrams 
would become immediately apparent in a graphical picture.

Therefore only planar diagrams survive, both in the Abelian and in the 
non-Abelian case. But, for planar diagrams, the only difference between the
two cases
is the appearance in the latter of the Casimir constant $C_F$. Hence
the Abelian exponentiation theorem continues to hold, leading to 
eq. (\ref{wlcpv}) (see eqs. (\ref{w2}), (\ref{abexp}) and (\ref{g2cpv})).

\section{}

\noindent
In this appendix we shall give the main sketch for the computation
of the independent diagrams $C_{(ij)(kl)}$ needed to 
derive the ${\cal O}(g^4)$ term in the perturbative expansion of the Wilson
loop ${\cal W}_\gamma^{ML}(L,T)$ in the causal formulation of the light-cone
gauge. As already explained in the main text, we can restrict ourselves to
the maximally non-Abelian diagrams, namely those providing a $C_FC_A$ factor.
Such diagrams are those in which the position of the
propagators is crossed, and there are 35 topologically inequivalent diagrams of
this type. However, thanks to the symmetry relations (\ref{sym}), the number of
independent diagrams to be evaluated is 19  (see eq. (\ref{relations})).

We first need the $E_{ij}(t,s)$ functions defined in eq. (\ref{e}) 
that are appropriate
to the present case: substituting the parametrization of the path (\ref{path})
and the propagator in the causal formulation (\ref{propml}) in eq. (\ref{e}),
we can derive all the functions  $E_{ij}(t,s)$. They are given by
\begin{eqnarray}
 E_{11}(t,s)\;=\;E_{33}(t,s)&=&-\frac{L^2}{4\pi\b^2}\nonumber\\
  E_{22}(t,s)\;=\;E_{44}(t,s)&=&\frac{L^2}{4\pi}\nonumber\\
  E_{12}(t,s)\;=\;E_{34}(t,s)&=&\frac{L^2}{4\pi\b}\;
                \frac{1-t+\b(1+s)}{1-t-\b(1+s)-i\e}\nonumber\\
  E_{23}(t,s)\;=\;E_{41}(t,s)&=&\frac{L^2}{4\pi\b}\;
                \frac{\b(1-t)-(1+s)}{\b(1-t)+1+s}\nonumber\\
  E_{13}(t,s)&=&\frac{L^2}{4\pi\b^2}\;
                \frac{t+s-2\b}{t+s+2\b+i\e}\nonumber\\
  E_{24}(t,s)&=&-\frac{L^2}{4\pi}\;
                \frac{\b t+\b s+2}{\b t+\b s-2}
\label{eml}
\end{eqnarray}
where,  $\beta=L/T$ and the symmetry relations  (\ref{sym}) have been
taken into account. The position  
(and the appearance) of poles in the above functions
 clearly
depends on the magnitude of $\beta$. Being interested in the large-$T$ limit,
we shall always consider the domain $\beta <1$. Consequently, the 
functions $ E_{23}(t,s)$, $E_{41}(t,s)$ and $ E_{24}(t,s)$ 
do not present poles: this is the reason why in
eq. (\ref{eml}) we omitted the  prescription for those functions as
irrelevant [to this purpose, remember that $s,t\in[-1,1]$, 
see eq. (\ref{path})].

The diagrams $C_{(ij)(kl)}$ are then defined in eq. (\ref{doppi}) as multiple
integrals of functions $E_{ij}(s,t)$. The notation is such that $C_{(ij)(kl)}$
denotes the diagram with two {\it crossed} propagators, the first joining the
segments $(\gamma_i , \gamma_j)$ and the second joining the segments
$(\gamma_k , \gamma_l)$. 
 Once one diagram $C_{(ij)(kl)}$ is evaluated, its value
has to be multiplied by a factor 8, which is the number of permutations 
of the indices $(ij)(kl)$ that maintains the position of the two
propagators crossed: 
this is a consequence of the first equation in (\ref{sym}). More
explicitly, this means 
\begin{equation}
C_{( i j )( k l )}\! =\!
C_{( ji )( kl )} \! = \!
C_{( ij )( lk )}\! = \!
C_{( ji )( lk )}\! = \!
C_{( kl )( ij )}\! = \!
C_{( lk )( ij )}\! = \!
C_{( kl )( ji )}\! = \!
C_{( lk )( ji )}\,.
\label{c}
\end{equation}
 To preserve
crossing,   the integration extrema have to be carefully chosen, and the
integration variables   $t,s,u,v$ have to be suitably nested.
Just as an example,  in the  diagram $C_{(11)(11)}$ the integration variables 
have to be such that $1>v>s>u>t>-1$ (see Fig. \ref{fig3}). Consequently, once
$t\in [-1,1]$, all the other integration extrema are fixed by the nesting, i.e.
$u\in [t,1]$, 
 $s\in [u,1]$,  $v\in [s,1]$. 

In the following calculation, we shall omit, for brevity, the factor $L^2/4\pi$,
which is common to all the propagators (\ref{eml}), defining ${\cal
E}_{(ij)(kl)}(t,s)=(4\pi/L^2)E_{(ij)(kl)}(t,s)$. The corresponding diagrams
will obviously rescale by a factor $(L^2/4\pi)^2$, and will be denoted by
${\cal C}_{(ij)(kl)}$, namely 
${\cal C}_{(ij)(kl)}=(4\pi/L^2)^2C_{(ij)(kl)}$.

Although in principle the evaluation of the 19 independent (rescaled) diagrams
is now clear, the practical calculation is rather cumbersome. We shall list here
the final results.   
\begin{eqnarray}
{\cal C}_{(\! 1 \! 1 \!)(\! 1 \! 1 \!)}&=&\!\!
\int_{-1}^1\dif t \int_t^1\dif u \int_u^1\dif s
                  \int_s^1\dif v\;{\cal E}_{11}(t,s)\;{\cal E}_{11}(u,v)
              =\frac{2}{3\b^4}\;\;\label{first}\\
&&{  }\nonumber\\
{\cal C}_{(\! 2 \! 2 \!)(\! 2 \! 2 \!)}&=&\!\!
\int_{-1}^1\dif t \int_t^1\dif u \int_u^1\dif s
                  \int_s^1\dif v\;{\cal E}_{22}(t,s)\;{\cal E}_{22}(u,v)
              =\frac{2}{3}\;\;\\
&&{  }\nonumber\\
{\cal C}_{(\! 1 \! 1 \!)(\! 1 \! 3 \!)}&=&\!\!
\int_{-1}^1\dif u \int_{-1}^1\dif v\;{\cal E}_{13}(u,v)
                  \int_{-1}^u\dif t \int_u^1\dif s\;{\cal E}_{11}(t,s)\nonumber\\
 &=&-\frac{8}{3\b^4}+\frac{64}{3\b^2}+\left(-\frac{16}{3\b^3}
   +\frac{16}{\b}-\frac{32}{3}\right)i\pi+\frac{64}{3}\ln(\b)+\nonumber\\
             &&+\left(\frac{16}{3\b^3}
   -\frac{16}{\b}-\frac{32}{3}\right)\ln(1+\b)+\left(-\frac{16}{3\b^3}
   +\frac{16}{\b}-\frac{32}{3}\right)\ln(1-\b)\;\;\\
&&{  }\nonumber\\
{\cal C}_{(\! 2 \! 2 \!)(\! 2 \! 4 \!)}&=&\!\!
\int_{-1}^1\dif u \int_{-1}^1\dif v\;{\cal E}_{24}(u,v)
                  \int_{-1}^u\dif t \int_u^1\dif s\;{\cal E}_{22}(t,s)\nonumber\\
             &=&\frac{64}{3\b^2}-\frac{8}{3}+\left(-\frac{32}{3\b^4}
   -\frac{16}{\b^3}+\frac{16}{3\b}\right)\ln(1+\b)
             +\left(-\frac{32}{3\b^4}
   +\frac{16}{\b^3}-\frac{16}{3\b}\right)\ln(1-\b)\;\;\\
&&{  }\nonumber\\
{\cal C}_{(\! 1 \! 1 \!)(\! 1 \! 2 \!)}&=&\!\!
\int_{-1}^1\dif u \int_{-1}^1\dif v\;{\cal E}_{12}(u,v)
                  \int_{-1}^u\dif t \int_u^1\dif s\;{\cal E}_{11}(t,s)\nonumber\\
            &=&-\frac{20}{3\b^2}+\frac{8}{\b}
               +\left(-\frac{32}{3\b}+8\right)i\pi
               +\left(\frac{32}{3\b}-8\right)\ln(\b)
            +\left(\frac{8}{3\b^4}
               -\frac{32}{3\b}+8\right)\ln(1-\b)\;\;\\
&&{  }\nonumber\\
{\cal C}_{(\! 2 \! 2 \!)(\! 2 \! 3 \!)}&=&\!\!
\int_{-1}^1\dif u \int_{-1}^1\dif v\;{\cal E}_{23}(u,v)
                  \int_{-1}^u\dif t \int_u^1\dif s\;{\cal E}_{22}(t,s)\nonumber\\
              &=&-\frac{8}{\b^3}-\frac{20}{3\b^2}-\frac{8}{3}\ln(\b)+
                 \left(\frac{8}{\b^4}+\frac{32}{3\b^3}+\frac{8}{3}
                  \right)\ln(1+\b)\;\;\\
&&{  }\nonumber\\
 {\cal C}_{(\! 1 \! 1 \!)(\! 1 \! 4 \!)}&=&\!\!
\int_{-1}^1\dif u \int_{-1}^1\dif v\;{\cal E}_{14}(u,v)
                      \int_{-1}^u\dif t \int_u^1\dif s\;{\cal E}_{11}(t,s)\nonumber\\
               &=&-\frac{20}{3\b^2}-\frac{8}{\b}-\left(\frac{32}{3\b}+8
                      \right)\ln(\b)+\left(\frac{8}{3\b^4}+\frac{32}{3\b}
                      +8\right)\ln(1+\b)\;\;\\
&&{  }\nonumber\\
{\cal C}_{(\! 2 \! 2 \!)(\! 1 \! 2 \!)}&=&\!\!
\int_{-1}^1\dif u \int_{-1}^1\dif v\;{\cal E}_{21}(u,v)
                  \int_{-1}^u\dif t \int_u^1\dif s\;{\cal E}_{22}(t,s)\nonumber\\
              &=&\frac{8}{\b^3}-\frac{20}{3\b^2}
                  +\frac{8}{3}\,i\pi-\frac{8}{3}\ln(\b)
                  +\left(\frac{8}{\b^4}-\frac{32}{3\b^3}
                  +\frac{8}{3}\right)\ln(1-\b)\;\;\\
&&{  }\nonumber\\
{\cal C}_{(\! 1 \! 2 \!)(\! 1 \! 3 \!)}&=&\!\!
\int_{-1}^1\dif u \int_{-1}^1\dif v\;{\cal E}_{12}(u,v)
                  \int_u^1\dif t \int_{-1}^1\dif s\;{\cal E}_{13}(t,s)\nonumber\\
  &=&\left(\frac{16\pi^2}{9}-\frac{8}{3}\right)\frac{1}{\b^3}
   +\left(\frac{16\pi^2}{3} -16\pi i\right)\frac{1}{\b^2}+\nonumber\\
       &&+\left(-\frac{16\pi^2}{3} +\frac{16\pi}{3}\,i
   \right)\frac{1}{\b}-16\pi i
       +\left(\frac{32}{3\b^2}+\frac{16}{3\b}
   +32+32\pi i\right)\ln(\b)+\nonumber\\
       &&+\left[\frac{16}{3\b^2}-\frac{32}{3\b} -16
   +\bigg(\frac{32\pi}{3\b^3}
   +\frac{32\pi}{\b^2}-\frac{64\pi}{3}\bigg)i\right]\ln(1+\b)
-32\ln^2(\b)+\nonumber\\
       &&+\left[-\frac{32}{3\b^4}+\frac{16}{\b^3}
   +\frac{16}{3\b^2}+\frac{16}{3\b}-16+\!\bigg(\!\!-\frac{64\pi}{3\b^3}
   +\frac{32\pi}{\b^2}-\frac{32\pi}{3}\bigg)i\right]\ln(1-\b)+\nonumber\\
       &&+\left[\frac{16}{3\b^3}+\frac{16}{\b^2}
   +\frac{32}{\b}+\frac{32}{3}\right]\ln^2(1+\b)
       +\left[-\frac{64}{3\b^3}+\frac{32}{\b^2}-\frac{32}{3}\right]
   \ln^2(1-\b)+\nonumber\\
       &&+\left[-\frac{32}{3\b^3}-\frac{32}{\b^2}-\frac{32}{\b}+
   \frac{32}{3}\right]\ln(\b)\ln(1+\b)+\nonumber\\
       &&+\left[-\frac{32}{3\b^3}-\frac{64}{\b^2}+\frac{32}{\b}
   +\frac{160}{3}\right]\ln(\b)\ln(1-\b)+\nonumber\\
       &&+\left[\frac{32}{\b^3}
   +\frac{32}{\b^2}-\frac{32}{\b}-32\right]\ln(1+\b)\ln(1-\b)+\nonumber\\
       &&+\left[-\frac{32}{3\b^3}-\frac{32}{\b^2}\right]\dil{\b}
       +\left[\frac{32}{3\b^3}+\frac{32}{\b^2}
   +\frac{32}{\b}+\frac{32}{3}\right]\dil{\frac{\b}{1+\b}}-\nonumber\\
       &&-\,\frac{32}{3}\;\dil{1-\frac{1}{\b}}
       +\left[-\frac{32}{3\b^3}-\frac{32}{\b^2}\right]
                                          \dil{\frac{1}{1+\b}}\;\; \\
&&{  }\nonumber\\
{\cal C}_{(\! 2 \! 3 \!)(\! 2 \! 4 \!)}&=&\!\!
\int_{-1}^1\dif u \int_{-1}^1\dif v\;{\cal E}_{23}(u,v)
  \int_u^1\dif t \int_{-1}^1\dif s\;{\cal E}_{24}(t,s)\nonumber\\[1mm]
     &=&-\,\frac{16\pi^2}{3\b^3}+\frac{16\pi^2}{3\b^2}+
  \left(\frac{8}{3}-\frac{16\pi^2}{9}\right)\frac{1}{\b}
     +\left[-\frac{64}{3\b^2}+\frac{16}{\b}+\frac{32}{3}\right]
  \ln(\b)+\nonumber\\
     &&+\left[-\frac{16}{\b^4}-\frac{16}{3\b^3}+\frac{16}{3\b^2}
  -\frac{16}{\b}-\frac{32}{3}\right]      \ln(1+\b)
+\left[-\frac{32}{\b^2}-\frac{64}{3\b}\right]    
  \ln(\b)\ln(1+\b)+\nonumber\\
  &&+\left[-\frac{16}{\b^4}+\frac{32}{3\b^3}+\frac{16}{3\b^2}
  \right]      \ln(1-\b)
     +\left[-\frac{16}{3\b^4}+\frac{32}{\b^2}+\frac{64}{3\b}
  \right]      \ln^2(1+\b)+\nonumber\\
     &&+\left[\frac{16}{3\b^4}-\frac{16}{\b^3}+\frac{16}{\b^2}
  -\frac{16}{3\b}\right]     \ln^2(1-\b)  
     +\left[-\frac{64}{\b^2}+\frac{128}{3\b}\right]
  \ln(\b)\ln(1-\b)+\nonumber\\
     &&+\left[-\frac{32}{\b^4}+\frac{32}{\b^3}+\frac{32}{\b^2}
  -\frac{32}{\b}\right]    \ln(1+\b)\ln(1-\b)+\nonumber\\
     &&+\left[-\frac{32}{3\b^4}+\frac{32}{\b^3}-\frac{32}{\b^2}
  +\frac{32}{3\b}\right]   \dil{1-\b}
     +\left[\frac{32}{\b^2}-\frac{32}{3\b}\right] 
  \dil{1-\frac{1}{\b}}+\nonumber\\
     &&+\,\frac{32}{3\b^4}\; \dil{\frac{1}{1+\b}}
     +\left[-\frac{32}{\b^2}+\frac{32}{3\b}\right]
  \dil{-\frac{1}{\b}}\;\;\\
&&{  }\nonumber\\
{\cal C}_{(\! 1 \! 3 \!)(\! 1 \! 4 \!)}&=&\!\!
\int_{-1}^1\dif u \int_{-1}^1\dif v\;{\cal E}_{14}(u,v)
  \int_{-1}^u\dif t \int_{-1}^1\dif s\;{\cal E}_{13}(t,s)\nonumber\\[1mm]
     &=&\left(\frac{8}{3}+\frac{16\pi^2}{9}\right)\frac{1}{\b^3}
  +\left(-\frac{16\pi^2}{3}+\frac{16\pi}{3}\,i\right)\frac{1}{\b^2}
     +\left(\frac{16\pi^2}{3}+\frac{32\pi}{3}\,i\right)\frac{1}{\b}
  -\frac{32\pi^2}{9}+\nonumber\\
     &&+\left[\frac{32}{3\b^2}-\frac{16}{3\b}+32+\frac{64\pi}{3}\,i
  \right]\ln(\b)-16\pi i  -\frac{80}{3}\ln^2(\b)+\nonumber\\
     &&+\left[-\frac{32}{3\b^4}\!-\!\frac{16}{\b^3}\!+\!\frac{16}{3\b^2}\!
  -\!\frac{16}{3\b}\!-\!16\!+\!\!\bigg(\!\!\!-\!\frac{32\pi}{\b^3}\!+\!\frac{32\pi}{\b^2}\!
  +\!\frac{32\pi}{\b}\!-\!32\pi\bigg)i\right]\!\ln(1+\b)+\nonumber\\
     &&+\left[\frac{16}{3\b^2}+\frac{32}{3\b}-16+\bigg(
  -\frac{32\pi}{3\b^3}+\frac{32\pi}{\b^2}-\frac{32\pi}{\b}
  +\frac{32\pi}{3}\bigg)i\right]\ln(1-\b)-\nonumber\\
      &&+\left[\frac{64}{3\b^3}+\frac{32}{\b^2}-\frac{16}{3}
  \right]\ln^2(1+\b)
     +\left[-\frac{32}{3\b^3}+\frac{32}{\b^2}-\frac{32}{\b}
  +\frac{32}{3}\right]\ln^2(1-\b)+\nonumber\\
     &&+\left[\frac{32}{3\b^3}-\frac{64}{\b^2}-\frac{32}{\b}
  +\frac{128}{3}\right]\ln(\b)\ln(1+\b)+\nonumber\\
     &&+\left[\frac{32}{3\b^3}-\frac{32}{\b^2}+\frac{32}{\b}
  +\frac{32}{3}\right]\ln(\b)\ln(1-\b)+\nonumber\\
     &&+\left[-\frac{32}{\b^3}+\frac{32}{\b^2}+\frac{32}{\b}
  -32\right]\ln(1+\b)\ln(1-\b)+\nonumber\\
     &&+\left[\frac{32}{3\b^3}-\frac{32}{\b^2}\right]\dil{-\b}
     +\,\frac{32}{3}\;\dil{\frac{\b}{1+\b}}+\nonumber\\
     &&+\left[-\frac{32}{3\b^3}+\frac{32}{\b^2}-\frac{32}{\b}
  +\frac{32}{3}\right]\dil{-\frac{\b}{1-\b}}
     +\left[-\frac{32}{3\b^3}+\frac{32}{\b^2}\right]\dil{1-\b}\;\;\\
&&{  }\nonumber\\
{\cal C}_{(\! 1 \! 2 \!)(\! 2 \! 4 \!)}&=&\!\!
\int_{-1}^1\dif u \int_{-1}^1\dif v\;{\cal E}_{21}(u,v)
  \int_{-1}^u\dif t \int_{-1}^1\dif s\;{\cal E}_{24}(t,s)\nonumber\\[1mm]
     &=&-\,\frac{16\pi^2}{3\b^3}+\left(-\frac{16\pi^2}{3}
  +\frac{64\pi}{3}\,i\right)\frac{1}{\b^2}
     +\left(-\frac{8}{3}-\frac{16\pi^2}{9}+16\pi i\right)
  \frac{1}{\b}-\frac{32\pi}{3}\,i+\nonumber\\
&&+\left[-\frac{16}{\b^4}-\frac{32}{3\b^3}+\frac{16}{3\b^2}
  +\bigg(-\frac{32\pi}{3\b^4}-\frac{32\pi}{\b^3}+\frac{64\pi}{3\b}
  \bigg)i\right]\ln(1+\b)+\nonumber\\
     &&+\left[-\frac{16}{\b^4}+\frac{16}{3\b^3}+\frac{16}{3\b^2}
  +\frac{16}{\b}-\frac{32}{3}\!+\!\bigg(\!\!-\frac{32\pi}{3\b^4}
  +\frac{32\pi}{\b^2}-\frac{64\pi}{3\b}\bigg)i\right]\ln(1-\b)+\nonumber\\
     &&+\left[\frac{32}{3\b^4}+\frac{32}{\b^3}+\frac{16}{\b^2}
  +\frac{16}{3\b}\right]\ln^2(1+\b)  +\left[-\frac{64}{3\b^2}
   -\frac{16}{\b}+\frac{32}{3}\right]
  \ln(\b)+\nonumber\\
     &&+\left[-\frac{32}{3\b^4}+\frac{32}{\b^2}-\frac{64}{3\b}
  \right]\ln^2(1-\b)
     +\left[-\frac{32}{\b^2}-\frac{32}{\b}\right]\ln(\b)\ln(1+\b)+\nonumber\\
    &&+\left[-\frac{32}{\b^2}+\frac{64}{3\b}\right]\ln(\b)\ln(1-\b)
     +\left[-\frac{32}{\b^4}-\frac{32}{\b^3}+\frac{32}{\b^2}
  +\frac{32}{\b}\right]\ln(1+\b)\ln(1-\b)+\nonumber\\
     &&+\left[\frac{32}{\b^2}+\frac{32}{3\b}\right]\dil{\b}
     +\left[-\frac{32}{\b^2}-\frac{32}{3\b}\right]
  \dil{\frac{\b}{1+\b}}-
     \frac{32}{3\b^4}\;\dil{1-\b}+\nonumber\\
     &&+\left[\frac{32}{3\b^4}+\frac{32}{\b^3}+\frac{32}{\b^2}
  +\frac{32}{3\b}\right]\dil{\frac{1}{1+\b}}\;\;\\
&&{  }\nonumber\\
{\cal C}_{(\! 1 \! 2 \!)(\! 1 \! 4 \!)}&=&\!\!
\int_{-1}^1\dif u \int_{-1}^1\dif v\;{\cal E}_{12}(u,v)
    \int_u^1\dif t \int_{-1}^1\dif s\;{\cal E}_{14}(t,s)\nonumber\\
     &=&\frac{4\pi^2}{9\b^4}+\left(\frac{8}{3}-\frac{16\pi^2}{3}
   +\frac{8\pi}{3}\,i\right)\frac{1}{\b^2}
     +\frac{16\pi}{\b}\,i+\frac{8\pi^2}{3}+\frac{28\pi}{3}\,i+
     \left[-\frac{16}{3\b^2}-\frac{56}{3}\right]\ln(\b)+\nonumber\\
     &&+\left[\frac{8}{3\b^3}-\frac{44}{3\b^2}-\frac{8}{\b}
   +\frac{28}{3}+\bigg(\frac{8\pi}{3\b^4}-\frac{32\pi}{\b^2}
   -\frac{64\pi}{3\b}+8\pi\bigg)i\right]\ln(1+\b)+\nonumber\\
     &&+\left[-\frac{8}{3\b^3}-\frac{44}{3\b^2}+\frac{8}{\b}
   +\frac{28}{3}+\bigg(\frac{8\pi}{3\b^4}-\frac{16\pi}{\b^2}
   +\frac{64\pi}{3\b}-8\pi\bigg)i\right]\ln(1-\b)+\nonumber\\
 &&+\left[-\frac{8}{3\b^4}+\frac{16}{\b^2}+\frac{128}{3\b}\right]
   \ln(\b)\ln(1+\b)+ 
\left[\frac{4}{3\b^4}-\frac{64}{3\b}-8\right]\ln^2(1+\b)+\nonumber\\
     &&+\left[-\frac{16}{3\b^4}+\frac{48}{\b^2}-\frac{128}{3\b}
   \right]\ln(\b)\ln(1-\b)+
\left[\frac{8}{3\b^4}-\frac{16}{\b^2}+\frac{64}{3\b}-8
   \right]\ln^2(1-\b)+\nonumber\\ 
     &&+\left[\frac{16}{\b^4}-\frac{32}{\b^2}+16\right]
   \ln(1+\b)\ln(1-\b)
     +\left[-\frac{8}{3\b^4}+\frac{32}{\b^2}\right]\dil{\b}+\nonumber\\
     &&+\left[\frac{8}{3\b^4}-\frac{16}{\b^2}-\frac{64}{3\b}-8\right]
   \dil{\frac{\b}{1+\b}}+
\left[-\frac{8}{3\b^4}+\frac{32}{\b^2}\right]
   \dil{\frac{1}{1+\b}}+\nonumber\\
     &&+\left[\frac{8}{3\b^4}-\frac{16}{\b^2}+\frac{64}{3\b}-8\right]
   \dil{-\frac{\b}{1-\b}} \;\;\\
&&{  }\nonumber\\
{\cal C}_{(\! 1 \! 2 \!)(\! 2 \! 3 \!)}&=&\!\!
\int_{-1}^1\dif u \int_{-1}^1\dif v\;{\cal E}_{23}(u,v)
                   \int_u^1\dif t \int_{-1}^1\dif s\;{\cal E}_{21}(t,s)\nonumber\\
    &=&\left(\frac{64\pi^2}{9}-8\pi i\right)\frac{1}{\b^3}
  +\left(\frac{8}{3}-\frac{52\pi}{3}\,i\right)\frac{1}{\b^2}
    -\frac{8\pi}{3\b}\,i+\frac{4\pi^2}{9}+
 \frac{40}{3}\ln^2(\b)+\nonumber\\
    &&+\left[\frac{28}{3\b^4}-\frac{8}{\b^3}-\frac{44}{3\b^2}
  +\frac{8}{3\b}+\bigg(\frac{8\pi}{\b^4}+\frac{64\pi}{3\b^3}
  +\frac{40\pi}{3}\bigg)i\right]\ln(1+\b)+\nonumber\\
    &&+\left[\frac{28}{3\b^4}+\frac{8}{\b^3}-\frac{44}{3\b^2}
  -\frac{8}{3\b}\right]\ln(1-\b)+
\left[\frac{104}{3\b^2}-\frac{40\pi}{3}\,i\right]\ln(\b)+\nonumber\\
   &&+\left[-\frac{8}{\b^4}-\frac{64}{3\b^3}+\frac{4}{3}\right]
  \ln^2(1+\b)
    +\left[-\frac{4}{\b^4}+\frac{32}{3\b^3}-\frac{8}{\b^2}
  +\frac{4}{3}\right]\ln^2(1-\b)+\nonumber\\
    &&+\left[\frac{16}{\b^2}-16\right]\ln(\b)\ln(1+\b)
    +\left[\frac{16}{\b^2}-16\right]\ln(\b)\ln(1-\b)+\nonumber\\
    &&+\left[\frac{16}{\b^4}-\frac{32}{\b^2}+16\right]
  \ln(1+\b)\ln(1-\b)
    +\left[-\frac{32}{\b^2}+\frac{8}{3}\right]\dil{\b}+\nonumber\\
   &&+\left[\frac{32}{\b^2}-\frac{8}{3}\right]\dil{\frac{\b}{1+\b}}
    +\left[\frac{8}{\b^4}-\frac{64}{3\b^3}+\frac{16}{\b^2}
  -\frac{8}{3}\right]\dil{1-\b}+\nonumber\\
    &&+\left[-\frac{8}{\b^4}-\frac{64}{3\b^3}-\frac{16}{\b^2}
  +\frac{8}{3}\right]\dil{\frac{1}{1+\b}}\;\;\\
&&{  }\nonumber\\
{\cal C}_{(\! 1 \! 2 \!)(\! 1 \! 2 \!)}&=&\!\!
\int_{-1}^1\dif u \int_{-1}^1\dif v\;{\cal E}_{12}(u,v)
                   \int_u^1\dif t \int_v^1\dif s\;{\cal E}_{12}(t,s)\nonumber\\
     &=&-\frac{4\pi^2}{3\b^4}+\left(-\frac{4}{3}+8\pi i
   \right)\frac{1}{\b^3}+\left(-\frac{4}{3}+4\pi i\right)
   \frac{1}{\b^2}
     -\left(\frac{4}{3}+\frac{8\pi i}{3}\right)\frac{1}{\b}
   -\frac{4\pi^2}{3}-\frac{28\pi}{3}\,i-\nonumber\\
     &&-\left[\frac{8}{\b^3}+\frac{4}{\b^2}-\frac{8}{3\b}
   -\frac{28}{3}+8\pi i\right]\ln(\b)+
     8\ln^2(\b)+
     \left[\frac{8}{\b^4}-\frac{16}{\b^2}+8\right]\ln^2(1-\b)+\nonumber\\
     &&+\left[-\frac{28}{3\b^4}+\frac{16}{3\b^3}+\frac{8}{\b^2}
   +\frac{16}{3\b}-\frac{28}{3}+\bigg(\frac{8\pi}{\b^4}
   -\frac{16\pi}{\b^2}+8\pi\bigg)i\right]\ln(1-\b)+\nonumber\\
     &&+\left[\frac{16}{\b^2}-16\right]\ln(\b)\ln(1-\b)+
     \frac{8}{\b^4}\;\dil{1-\b}+
     8\,\dil{1-\frac{1}{\b}}\;\;\\
&&{  }\nonumber\\
{\cal C}_{(\! 2 \! 3 \!)(\! 2 \! 3 \!)}&=&\!\!
\int_{-1}^1\dif u \int_{-1}^1\dif v\;{\cal E}_{23}(u,v)
                   \int_u^1\dif t \int_v^1\dif s\;{\cal E}_{23}(t,s)\nonumber\\
     &=&\frac{4\pi^2}{3\b^4}+\frac{4}{3\b^3}-\frac{4}{3\b^2}
   +\frac{4}{3\b}+\frac{4\pi^2}{3}
     +\left[\frac{8}{\b^3}-\frac{4}{\b^2}-\frac{8}{3\b}+\frac{28}{3}
   \right]\ln(\b)+\nonumber\\
     &&+\left[-\frac{28}{3\b^4}-\frac{16}{3\b^3}+\frac{8}{\b^2}
   -\frac{16}{3\b}-\frac{28}{3}\right]\ln(1+\b)+
     4\ln^2(\b)+\nonumber\\
     &&+\left[\frac{4}{\b^4}-\frac{16}{\b^2}+4\right]\ln^2(1+\b)
     +\left[\frac{16}{\b^2}-8\right]\ln(\b)\ln(1+\b)-\nonumber\\
     &&-\,8\,\dil{\frac{\b}{1+\b}}
     -\frac{8}{\b^4}\;\dil{\frac{1}{1+\b}}\;\;\label{intermediate}\\
&&{  }\nonumber\\
{\cal C}_{(\! 1 \! 3 \!)(\! 1 \! 3 \!)}&=&\!\!
\int_{-1}^1\dif u \int_{-1}^1\dif v\;{\cal E}_{13}(u,v)
                   \int_u^1\dif t \int_v^1\dif s\;{\cal E}_{13}(t,s)\nonumber\\
     &=&\frac{4}{\b^4}+\frac{32\pi}{3\b^3}\,i
   -\frac{32}{3\b^2}-\frac{64\pi}{\b}\,i+\frac{160\pi}{3}\,i-
     \left[\frac{320}{3}+64\pi i\right]\ln(\b)+\nonumber\\
     &&+\left[-\frac{32}{3\b^3}-\frac{64\pi}{\b^2}\,i+\frac{64}{\b}
   +\frac{160}{3}+64\pi i\right]\ln(1+\b)+
 64\ln^2(\b)+\nonumber\\
     &&+\left[\frac{32}{3\b^3}-\frac{64}{\b}+\frac{160}{3}
   \right]\ln(1-\b)
    +\left[\frac{64}{\b^2}-64\right]\ln(\b)\ln(1+\b)-\nonumber\\
     &&-\,64\ln(\b)\ln(1-\b)+
     \left[-\frac{64}{\b^2}+64\right]\ln(1+\b)\ln(1-\b)+\nonumber\\
     &&+\,\frac{64}{\b^2}\;\dil{-\b}-
    \,\frac{64}{\b^2}\;\dil{1-\b}\;\;\\
&&{  }\nonumber\\
{\cal C}_{(\! 2 \! 4 \!)(\! 2 \! 4 \!)}&=&\!\!
\int_{-1}^1\dif u \int_{-1}^1\dif v\;{\cal E}_{24}(u,v)
                   \int_u^1\dif t \int_v^1\dif s\;{\cal E}_{24}(t,s)\nonumber\\
     &=&-\,\frac{32}{3\b^2}+4 +\frac{32}{\b^2}\ln^2(1+\b)
   +\left[\frac{64}{\b^4}-\frac{64}{\b^2}\right]\ln(1+\b)\ln(1-\b)+\nonumber\\
     &&+\left[\frac{160}{3\b^4}+\frac{64}{\b^3}-\frac{32}{3\b}\right]
       \ln(1+\b)+
     \left[\frac{160}{3\b^4}-\frac{64}{\b^3}+\frac{32}{3\b}\right]
       \ln(1-\b)-\nonumber\\
     &&-\,\frac{64}{\b^2}\;\dil{\b}+
     \frac{64}{\b^2}\;\dil{\frac{\b}{1+\b}}\;\;\\
&&{  }\nonumber\\
{\cal C}_{(\! 1 \! 3 \!)(\! 2 \! 4 \!)}&=&\!\!
\int_{-1}^1\dif t \int_{-1}^1\dif s\;{\cal E}_{13}(t,s)
                   \int_{-1}^1\dif u \int_{-1}^1\dif v\;{\cal E}_{24}(u,v)\nonumber\\
      &=&-\,\frac{16}{\b^2}-\frac{32\pi}{\b}\,i+32\pi i-
      \,64\ln(\b)+\nonumber\\
      &&+\left[\frac{32}{\b^4}+\frac{32}{\b^3}+\frac{32}{\b}
   +32+\bigg(\frac{64\pi}{\b^3}-\frac{64\pi}{\b}\bigg)i\right]\ln(1+\b)+\nonumber\\
      &&+\left[\frac{32}{\b^4}-\frac{32}{\b^3}-\frac{32}{\b}
   +32+\bigg(\frac{64\pi}{\b^3}-\frac{128\pi}{\b^2}+\frac{64\pi}{\b}\bigg)i
   \right]\ln(1-\b)+\nonumber\\
      &&+\left[-\frac{64}{\b^3}-\frac{128}{\b^2}-\frac{64}{\b}\right]
   \ln^2(1+\b)
      +\left[\frac{64}{\b^3}-\frac{128}{\b^2}+\frac{64}{\b}\right]
   \ln^2(1-\b)+\nonumber\\
      &&+\left[\frac{128}{\b^2}+\frac{128}{\b}\right]\ln(\b)\ln(1+\b)+
    \left[\frac{128}{\b^2}-\frac{128}{\b}\right]\ln(\b)\ln(1-\b)\;\;\ .
\label{last}
\end{eqnarray} 
Some technical details on the dilogarithm function ${\rm
Li}_2(z)$ and on its analytic continuations are in order.
As is well known, ${\rm Li}_2(z)$ can be defined through its integral
representation
\begin{equation}
\dil{z}=\int_z^0\frac{\ln(1-\zeta)}{\zeta}\;\diff\zeta\;\;,
\label{dil}
\end{equation}
where the path joining $z$ and $0$ is arbitrary, provided it does not intersect 
the half-line $]1,+\infty[$~, which is the branch-cut of the integrand function.
On its branch-point the dilogarithm is finite, and takes the value
$\dil{1}=\pi^2/6$. 

If $\beta <1$, the calculation of the above diagrams 
involves dilogarithmic functions, with arguments bounded by the region $-\infty
<{\rm Re}\, z<1$. Eventually, we shall be 
interested in taking the limit $\beta \to
0$ (i.e. large $T$). The arguments of the dilogarithms arising from a first 
integrations of eqs. (\ref{first})--(\ref{last}) 
can tend to $0$, $1$ and $-\infty$
as $\beta\to 0$.
 On the other hand, the simplest expansion of the dilogarithm 
is around the point $z=0$, where a simple series representation holds
\begin{equation}
\dil{z}=\sum_{k=1}^{\infty}\frac{z^k}{k^2} \ ,\qquad \ \ |z|<1\ \ .
\label{dil2}
\end{equation}
Consequently, we need analytic continuation to convert dilogarithms with
arguments tending to $1$ and $-\infty$ into dilogarithms with arguments tending
to $0$, for $\beta\to 0$. These are given by
\begin{eqnarray}
\dil{-z}&=&-\frac{\pi^2}{6}+\frac{1}{2}\ln^2(1+z)-\ln(z)\ln(1+z)
           +\dil{\frac{1}{1+z}}\ \,\ ,\label{ac1}\\
\dil{z}&=&\frac{\pi^2}{6}-\ln(z)\ln(1-z)-\dil{1-z}.
\label{ac2}
\end{eqnarray}
In our specific case, the following formulas, which can be derived from eqs.
(\ref{ac1}), (\ref{ac2}) have been repeatedly used to get the final result from
eqs.(\ref{first})--(\ref{last}):
\begin{eqnarray}
\dil{-\frac{1}{\b}}&=&-\frac{\pi^2}{6}
   +\frac{1}{2}\ln^2(1+\b)-\frac{1}{2}
   \ln^2(\b)+\dil{\frac{\b}{1+\b}}\;\;, \label{ac3}\\
 \dil{1-\frac{1}{\b}}&=&-\frac{\pi^2}{6}+\frac{1}{2}\ln^2\left(
 \frac{1}{\b}\right)-\ln\left(\frac{1-\b}{\b}\right)\ln\left(
 \frac{1}{\b}\right)+\dil{\b}\;\;,\label{ac4}\\
\dil{1-\b}&=&\frac{\pi^2}{6}-\ln(1-\b)\ln(\b)-\dil{\b}\;\;,\label{ac5}\\[1mm]
  \dil{\frac{1}{1+\b}}&=&\frac{\pi^2}{6}-\ln\left(\frac{\b}{1+\b}
    \right)\ln\left(\frac{1}{1+\b}\right)-\dil{\frac{\b}{1+\b}}\;\;.\label{ac6}
\end{eqnarray}
In this way, all the dilogarithms in eqs. (\ref{first})--(\ref{last}) are ready
to be  easily expanded in power series of $\beta$ as 
\begin{eqnarray}
\dil{\b}&=&+\b+\frac{1}{4}\b^2+\;\frac{1}{9}\b^3\,+{\cal O}(\b^4)\ ,
\label{dilex1}\\[2mm]
  \dil{-\b}&=&-\b+\frac{1}{4}\b^2-\;\frac{1}{9}\b^3\,+{\cal O}(\b^4)\
,\label{dilex2}\\[1.5mm]
  \dil{\frac{\b}{1+\b}}&=&+\b-\frac{3}{4}\b^2+\frac{11}{18}\b^3
                          +{\cal O}(\b^4)\ ,\label{dilex3}\\
  \dil{-\frac{\b}{1-\b}}&=&-\b-\frac{3}{4}\b^2-\frac{11}{18}\b^3
 +{\cal O}(\b^4) \ .\label{dilex4}
\end{eqnarray}
\noindent
To get the final result for the maximally non-Abelian ${\cal O}(g^4)$ terms in
the causal formulation, we have 
\begin{enumerate}
\item to sum all the results (\ref{first})--(\ref{last}), taking into
account the analytic continuations eqs. (\ref{ac3})--(\ref{ac6}); 
the integrals from eqs.
(\ref{first}) to (\ref{intermediate}) have to be multiplied by an extra 
 factor 2 to take into account the 16 relations (\ref{relations});
\item to multiply the result by  $L^4/(4\pi)^2$ to take into account the rescaling
from ${\cal C}_{(ij)(kl)}$ to ${C}_{(ij)(kl)}$;
\item to multiply by a factor 8 to take into account permutations of indices as in
eq. (\ref{c});
\item to multiply by a factor $-C_AC_F/16$ as shown in eq. (\ref{doppi}).
\end{enumerate}
\noindent 
Following the above points (1) to (4) one arrives at eq. (\ref{wmlg4}), whereas
substituting eqs. (\ref{dilex1})--(\ref{dilex4}) in (\ref{wmlg4}) one can easily
get the large-$T$ behaviour of the result, namely eq. (\ref{limit}).

\vfill\eject

\begin{figure}
\caption{Parametrization of the closed  rectangular loop $\gamma$ in four
segments $\gamma_i$.}
\label{fig1}
\end{figure}
\begin{figure}
\caption{Example of non-crossed and crossed diagrams.}
\label{fig2}
\end{figure}
\begin{figure}
\caption{Examples of crossed diagrams; they are labelled as $C_{(11)(11)}$,
$C_{(23)(34)}$ and $C_{(13)(14)}$.}
\label{fig3}
\end{figure}
\begin{figure}
\caption{The three crossed diagrams that are unrelated to other diagrams through
eq. (\ref{sym}); they are  $C_{(13)(13)}$,
$C_{(24)(24)}$ and $C_{(13)(24)}$. }
\label{fig4}
\end{figure}
\begin{figure}
\caption{Examples of spider diagrams.}
\label{fig5}
\end{figure}

\vfill
\eject

\begin{picture}(400,820)(33,0)
\leavevmode
\epsfbox{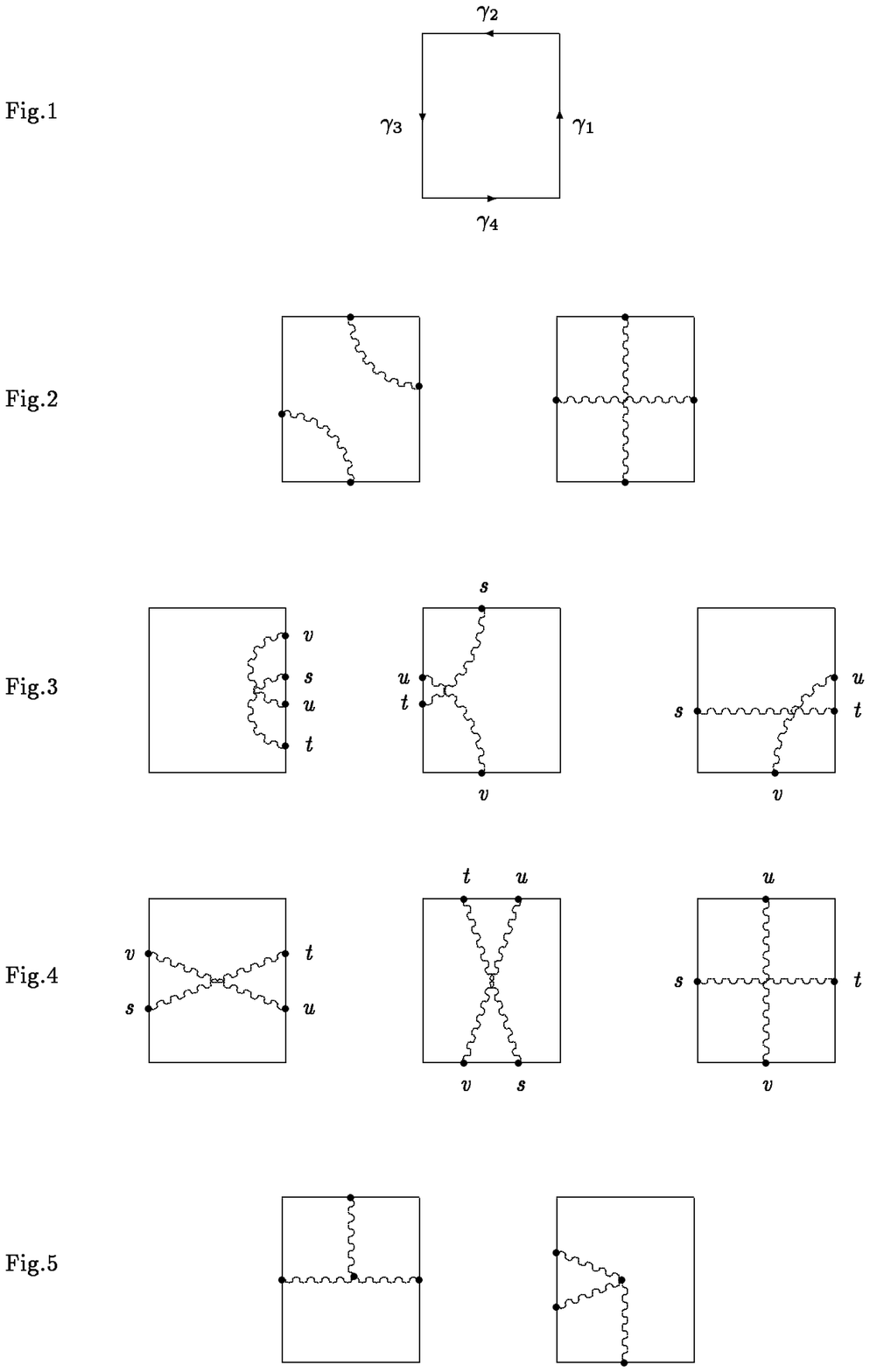}
\end{picture}


\begin{references}
\bibitem{nota1}{Usually the potential is defined in the corresponding
Euclidean formulation. Here a Minkowskian treatment is forced by the
light-cone gauge choice. The two treatments should be related by 
analytic continuation.}
\end{references}
\end{document}